\documentclass[prd,superscriptaddress,nofootinbib,showpacs,showkeys,preprint]{revtex4}
\usepackage[colorlinks]{hyperref}
\usepackage{amsmath,amssymb}

\def\beq{\begin{equation}}
\def\eeq{\end{equation}}
\def\bea{\begin{eqnarray}}
\def\eea{\end{eqnarray}}

\renewcommand{\thefootnote}

\begin{document}
\title{Generalized quantum electrodynamics in  the framework of Generalized BRST transformation}
\author{Krishnanand Kr. Mishra, Bhabani Prasad Mandal}
\email{mishra.krishna93@gmail.com;bhabani.mandal@gmail.com}
\affiliation{Department of Physics,  Institute of science, Banaras Hindu University, 
Varanasi  - 221005, India}

\begin{abstract}
 Podolsky's electromagnetism which explains some of unresolved problems of usual QED has been investigated in the framework of finite field 
dependent (FFBRST) BRST transformation. In this generalized QED (GQED),  BRST invariant effective theories are written using generalized Lorenz gauge, $(1+\frac{\Box }{m_p^2})\partial^\mu A_\mu=0$  and no mixing gauge condition $(1+\frac{\Box }{m_p^2})^\frac{1}{2} \ \partial^\mu A_\mu=0$, it  contains higher order derivative terms. No Mixing gauge is free from UV divergences in the radiative correction and easier to handle. We construct appropriate FFBRST transformation to obtain generating functional in no mixing gauge from that of in Lorenz gauge.  We show that all these BRST invariant effective theories in GQED are basically same and are inter-connected  with appropriately constructed FFBRST transformations.
 \end{abstract}
 \maketitle

 \section {Introduction}
 A linear and local extension of Maxwell theory for electromagnetism was proposed by Podolsky in 1940's \cite{1} to overcome certain divergence difficulties known at that time.This theory respects Poincare symmetry and U(1) gauge symmetry,  contains higher order derivative terms and is characterized by a free parameter, known as Podolsky mass $m_p$. However, such theory initially failed to achieve its goal due to lack of appropriate gauge fixing in the theory.
In spite of initial failures of this generalized QED (GQED) formulation, it has been used to explain some of the unresolved problems in usual QED \cite{3}.  Electrostatic potential diverges over punctual electric charges at the classical level, but is finite everywhere in GQED formulation \cite{1,2}.
There are difficulties associted with the concepts of electromagnetic momenta and energy in Maxwell theory,  which are not taken care of at the quantum level. This problem is referred in  literature as $"\frac{4}{3}$ problem in classical electrodynamics" \cite{4}. Further, GQED provides a richer theoretical foundation compare to usual QED and introduces a new length scale in theory $l_p=\frac {1}{\mid m_p \mid }$. Such a higher derivative theory is local, stable and unitary \cite{6}. Unlike usual QED,  where non-covariant Coulomb gauge is used to work with true degree of freedom of the theory, a kind of generalized Lorenz gauge condition $[1+\frac{\Box}{m_p^2}]^\varepsilon \partial_\mu A^\mu =0 $ is generally used in GQED. Generalized Lorenz guage with $\varepsilon =0$ reduces to usual Lorenz gauge and on the other hand  for $\varepsilon =1/2$ it corresponds to no mixing gauge. No Mixing gauge is generally applied to calculate Casimir effect in Podolsky QED and easier to handle and no UV divergence appears in the radiative correction  \cite{6}. BRST invariant effective theories are constructed in these gauges.

In this present work we investigate GQED in the framework of FFBRST formalism \cite{7} which is capable of connecting various effective theories through field transformation. FFBRST transformations are the generalization of usual BRST transformation, where the usual infinitesimal anti-commuting global transformation parameter is replaced by a finite field dependent but anti-commuting parameter. Such a finite generalization of BRST transformation protects nilpotency and retains the symmetry of gauge field actions. Due to the finiteness of the parameter the path integral measure transform non trivially under FFBRST. Under certain condition the Jacobian of the path integral measure is expressed as local functional of the fields and its derivatives  \cite{7}. Jacobian contribution is adjusted by choosing appropriate parameter in FFBRST transformation. Thus FFBRST with appropriate parameter can relate the generating functional of different effective theories. In virtue of this remarkable property, FFBRST transformation have been investigated extensively and have found many applications  in various gauge field theoretic models. \cite{8,9,10,11,12,13,14,15,16,17,171,18}

  In the present work we investigate various effective actions in GQED in the framework of FFBRST transformation. We construct appropriate FFBRST finite parameter which interrelates the generating functionals with the effective actions in Lorenz gauge,  no mixing gauge and generalized Lorenz gauge. This establishes that various effective actions in Podolsky's electromagnetism are interrelated through field transformations. In particular we obtain generating functional corresponding to no mixing gauge, which is free from UV divergence from that of in Lorenz gauge using appropriately constructed FFBRST.

          Now we present the plan of paper. In the next section we briefly review GQED or Podolsky's theory of electromagnetism. In Section III, we outline the technique of FFBRST transformation with an example. Interrelation of various effective theories through FFBRST are presented in Section  IV. Final section is kept for discussion and conclusion. 
 
\section {\bf Generalized QED(GQED)}
The Lagrangian density is written in generalized QED with second order derivative term as-
\begin{eqnarray}
\mathcal{L} =-\frac{1}{4} F_{\mu\nu} F^{\mu\nu}+\frac{1}{2m_p^2}\partial_ {\mu} F^{\mu\lambda} \partial_{\lambda} {F_{\nu}}^{\lambda}.
\end{eqnarray}
where the field strength tensor $F_{\mu\nu}$ has the usual structure  $F_{\mu\nu}=\partial_{\mu} A_{\nu}-\partial_\mu A_\nu$. The free parameter $m_p$ is dimensionful and usual Maxwell theory is recovered in the limit $m_p {\rightarrow}\infty $\\
The Euler-Lagrange equation of motion for this theory is
\begin{eqnarray} 
[\Box+m_p^2]\partial_\mu F^{\mu\lambda}=0
\end{eqnarray}
where $\Box=\partial_\mu\partial^\mu$ is d'Almbert in Minkowski space.
The equation of motion is different from that in Maxwell theory.  Hence there is possibility of containing new physics in this extended theory of electromagnetism. One can impose generalized Lorenz condition $[\Box+m_p^2]\partial_\mu A^\mu=0$
in the Podolsky field to simplify the equation of motion in Eq. (2) to obtain
\begin{eqnarray}
 [\Box+m_p^2]\Box A^\mu=0
\end{eqnarray}
The possible solution for this equation is constructed  by denoting 
\begin{eqnarray}
A^\mu =A_M^\mu + A_P^\mu.
\end{eqnarray}
where $A_M^\mu$is Maxwell field satisfying the condition
\begin{eqnarray}
\Box A_M^\mu= 0
\end{eqnarray} 
 and $A_P^\mu$ is Proca field satisfying the condition
\begin{eqnarray}
[\Box +m_p^2]A_P^\mu= 0
\end{eqnarray} 
The path integral formulation was carried out systematically for this formulation and the generating functional is defined as, 
\begin{eqnarray}
Z=\int{{\mathcal{D}}A {\mathcal{D}}B{\mathcal{D}}c{\mathcal{D}}\bar c} \ e^{i\int{d^4x}{\mathcal{L}}_{eff}}
\end{eqnarray}
where
 \begin{eqnarray}            
{\mathcal{L}}_{eff}=-\frac{1}{4} F_{\mu\nu} F^{\mu\nu}+\frac{1}{{2}{m_p}^{2}}\partial_{\mu} F^{\mu\lambda} \partial_\theta F_{\theta}^{\lambda} +
B{G[A]}+\frac{\xi}{2} B^2 
+\bar c(\frac \Box {{m_p}^{2}}+1)^\varepsilon \Box c
\end{eqnarray}
where $ G[A]$ is gauge fixing term corresponds to generalized Lorenz gauge. B is auxiliary field, c and $\bar c $ are ghost and anti-ghost fields respectively. \\
Above Lagrangian density is then expressed as,
\begin{eqnarray}
{\mathcal{L}}_{eff}={\mathcal{L}}_{o}+\delta_{BRST}[\bar c (\frac {B}{2}-(\frac{\Box }{{m_p}^2}+1)^\varepsilon )\partial_\mu{A}^\mu]
\end{eqnarray}
The effective action is invariant under the BRST transformation is given by
\begin{eqnarray}
\delta A_\mu &= &-{\partial_\mu c}\delta\omega \nonumber\\
\delta c &=& 0\nonumber\\
\delta{\bar c} &=& B \delta \omega \nonumber\\
\delta B &=& 0\nonumber\\
\end{eqnarray}
where $\delta\omega$ is global, anticommuting,infinitesimal parameter.

\section {Finite field dependent BRST transformation} 
Now we briefly outline the procedure to generalize  BRST transformation in QED which is described by the effective action in Lorenz gauge
in terms of auxiliary field  B as,
\begin{eqnarray}
 S_{eff}^{L}=\int{d^4x}[-\frac{1}{4} F_{\mu\nu} F^{\mu\nu}+\frac{\lambda}{2} B^2+B(\partial_\mu A^\mu )- \bar c \partial^\mu \partial_\mu c]
\end{eqnarray} 
This theory is invariant under  BRST transformation given in Eq. (10)
which we generically denote as 
\begin{eqnarray}
{\delta \phi}={\delta_b}{\phi}{\delta \omega} 
\end{eqnarray}
where ${\phi}=\{A,c,\bar c, B \}$,  a generic notation for all the fields.
We start, with making the infinitesimal global parameter  ${\delta \omega}=(\Theta^{'}(\phi(x,\kappa))d\kappa)$ field dependent by introducing a numerical parameter  $\kappa \ (0\leq \kappa \leq 1)$ and making all the fields $\kappa $ dependent such that  $\phi  (x,\kappa =0)=\phi(x)$  and $\phi  (x,\kappa =1)=\phi^{'}(x)$ the transformed field.\\ 
The BRST transformation in Eq.(12) is then written as
 \begin{eqnarray}
 d\phi =\delta _b[\phi (x,\kappa )]\Theta ^{'}(\phi (x,\kappa ))d\kappa
\end{eqnarray} 
where $\Theta^{'}$ is a field dependent anti-commuting parameter and $\delta_b\phi (x,\kappa )$  BRST variation
for the corresponding field as in Eq.  (10). The FFBRST is then constructed by integrating Eq. (13) from $\kappa =0$ to $\kappa =1$ as \cite{7}. 
\begin {eqnarray}
\phi^{'}=\phi (x,\kappa =1)=\phi (x,\kappa =0)+\delta_b[\phi (x,\kappa =0)]\Theta [\phi (x)]
\end{eqnarray}
where $\Theta [\phi (x)]=\int d\kappa ^{'}\Theta ^{'} [\phi (x,\kappa )]$.\\
Like usual BRST transformation, FFBRST transformation leaves the effective actions in Eq. (11) invariant. However, since the transformation parameter is field dependent  and finite in nature, 
FFBRST transformation  does not leave the path integral measure, $ D \phi=\prod_{\mu,x}dA_\mu(x)dc(x)d\bar c(x)dB$ invariant. It produces a non trivial Jacobian factor J, which further can be cast as a local functional of fields,  $e^{iS_J}$ (where the $S_J$ is the action representing the Jacobian factor J) 
if the following condition is met \cite{7} 
\begin{eqnarray}
\int D\phi (x,\kappa )[\frac{1}{J}\frac{dJ}{d\kappa }-i\frac{dS_J}{d\kappa}] e^{i(S_J+S_{eff})}=0
\end{eqnarray}
Thus the procedure of FFBRST may be summarised as \\ (i) calculate the infinitesimal change in Jacobian $\frac{1}{J}\frac{dJ}{d\kappa }{d\kappa }$ using 
\begin{eqnarray}
 \frac{J(\kappa )}{J(\kappa +d\kappa) }=1-\frac{1}{J(\kappa )}\frac{dJ(\kappa )}{d\kappa }d\kappa =\pm \sum_\phi  \frac{\delta \phi (\kappa +d\kappa )}{\delta \phi(\kappa)}
\end{eqnarray} 
for infinitesimal BRST transformation, + or - sign is for Bosonic or Fermionic nature of the field $\phi$ respectively\\ (ii) to find Jacobian contribution make an ansatz for $S_J$\\(iii) then prove the Eq. (15 ) for this ansatz  and finally \\(iv) replace J $(\kappa )$ by $e^{iS_J} $ in the generating functional
\begin{eqnarray}
 J(\kappa )\longrightarrow e^{iS_J[\phi (x,\kappa)]}
\end{eqnarray}

Thus FFBRST transformation changes the effective action inside the path integral through the Jacobian as
\begin{eqnarray}
S_{eff}^{'}=S_J+S_{eff}
\end{eqnarray}

Hence
\begin{eqnarray} 
Z=\int D\phi (x) e^{iS_{eff}[\phi ( x)]} \  \  \stackrel{FFBRST}{\longrightarrow} \ \ \int D\phi e^{iS_{eff}(\phi)+iS_J(\phi)} = Z^\prime
\end{eqnarray}               

 Thus FFBRST transformation connects two two generating functionals corresponding  to two  BRST invariant effective actions.\\
To illustrate this we consider a definite example by connecting t'Hooft-Veltmann effective theory to Lorenz theory. The effective action in t'Hooft -Veltmann gauge is described as
\begin{eqnarray}
S_{eff}^t=\int{d^4x}[-\frac{1}{4} F_{\mu\nu} F^{\mu\nu}+ \frac{\lambda}{2} B^2 -
B [\partial_\mu A^\mu +g{A_\mu}{A^\mu}]-\bar c[ \Box  +2 g{A^\mu}{\partial_\mu}]c]
\end{eqnarray}
This effective action is invariant under the following usual BRST transformation in Eq.(10)
Now we construct a FFBRST transformation with finite field dependent parameter $\Theta=\int{d\kappa }\Theta^{'}[\phi(x,\kappa)]$ where\\
\begin{eqnarray}
\Theta ^{'}=i\int {d^4x}{\bar c}[ \gamma_ 1 \lambda B+ \gamma _2  \partial_\mu A^\mu - \gamma_ 3 g A_\mu  A^\mu ]
\end{eqnarray}
with $\gamma_1 ,\gamma_2 ,\gamma_3$ are arbitary constants.\\
Furthermore,  the  infinitesimal change in Jacobian due to this, FFBRST is calculated using Eq.(16) as follows \\
\begin{eqnarray}
\frac{1}{J(\kappa )}\frac{dJ}{dk}&=&\int d^4x[({s_b}{A_\mu})\frac{\delta{\Theta}^{'}}{\delta{A_\mu}}-({s_b}{c}) \frac{\delta{\Theta}^{'}}{ \delta{c}}-
(s_b \bar c) \frac{\delta{\Theta}^{'}}{\delta{\bar c}}+({s_b} B) \frac{\delta{\Theta}^{'}}{\delta B}]\nonumber\\
&=&\int{d^4x}[{\gamma_ 2} \bar c \Box c+
2 \gamma_ 3 g {\bar c}{A^\mu}{\partial_\mu}{c}-{\gamma_ 1}{\lambda }{B}^{2}-\gamma_ 2 B(\partial_\mu A^\mu )+
\gamma _3 B gA_\mu A^\mu]
\end{eqnarray}
We make an ansatz for local functional $S_J$ which appears in the exponential of Jacobian as,
\begin{eqnarray}
S_{J} = [ \xi _1 {B^2}+ \xi _2 B(\partial_\mu A^\mu)+ \xi _3  B(\partial_\mu A ^\mu +gA_\mu A^\mu)
+ &\xi_ 4  {\bar c} \Box c+ \xi_ 5 {\bar c} ({\Box +2gA^\mu }{\partial_\mu})c]
\end{eqnarray}
where ${\xi_i}$ depends on $\kappa $, satisfying ${\xi_i}(\kappa =0)=0$.
\begin{eqnarray}
\frac{dS_J}{dk}&=&(\xi_ 1^{'}B^{2}+\xi_ 2^{'}{B}(\partial_\mu A^\mu) + \xi_ 3^{'}(\partial_\mu A^\mu) +\xi_ 3^{'}{B}{gA_\mu A^\mu}
+\xi_ 4^{'}\bar c \Box  c+\xi_ 5^{'}\bar c \Box c+\xi_ 5^{'}\bar c 2g{A^\mu} \partial_\mu  c\nonumber\\
&-&\xi_ 2 B \Box c \theta^{'}
-\xi_ 3 B \Box c \theta^{'}-2 {\xi_ 3} B g{A^\mu}{\partial^\mu}{c}\theta^{'}
+{\xi_ 4}{B}{\theta}^{'}{\Box}{c}+{\xi_ 5}{B}{\theta}^{'}{\Box }{c}\nonumber\\
&+&{2g}{\xi_ 5}{B}{\theta}^{'}{A^\mu}{\partial_\mu}{c}-{2}{g}{\xi_ 5}{\bar c}{\Box}{c}{\theta^{'}})
\end{eqnarray}
where primes denotes derivatives with respect to $\kappa $.
Now we substitute  Eq.(22) and Eq.(24) in Eq.(15) to obtain 
\begin{eqnarray}
\begin{split}
\int {d^4x}[-(\xi _1^{'}+\gamma _1 \lambda ) B^{2}-(\xi_ 2^{'}+\xi_ 3^{'}+\gamma _2)B(\partial_\mu A^\mu)-(\xi_ 3^{'}-\gamma _3 ){B gA_\mu A^\mu}\\
 -(\xi_ 4^{'}+\xi_ 5^{'}+\gamma _2){\bar c \Box  c}-(\xi_ 5^{'}-\gamma _3 ){2g\bar c A^\mu \partial_\mu c}+({\xi_ 2+\xi_ 3-\xi_ 4-\xi_ 5}){B}{\Box}{c}{\Theta^{'}}\\
-{2g}{\xi_ 5}{B}{\Theta^{'}}A^{\mu}{\partial_\mu}{c}+{2}{\xi_ 3}{g}{B}{A^\mu}{\partial_\mu}{c} \Theta^{'}+2g\xi_ 5 \bar c \Box c \Theta^{'}] e^{i(S_{eff}+S_1)}=0\\
\end{split}
\end{eqnarray}
 Now equating the coefficient of various terms,  we get the following differential equation for coefficient $\xi_{i}$.
\begin{eqnarray}
\xi _1^{'}+\gamma _1 \lambda =0 \nonumber\\
\xi_ 2^{'}+\xi_ 3^{'}+\gamma _2=0\nonumber\\
\xi_ 3^{'}-\gamma _3 =0\nonumber\\
\xi_ 4^{'}+\xi_ 5^{'}-\gamma _2=0\nonumber\\
\xi_ 5^{'}- \gamma_ 3  = 0\nonumber\\
\xi_ 2+\xi_ 3 +\xi_ 4+\xi_ 5=0\nonumber\\
\xi_ 3- \xi_ 5=0\nonumber\\
\end{eqnarray}
The solutions of the differential equations,  fulfilling the boundary conditions ${\xi} _{i}(\kappa =0)=0$,  are 
\begin{eqnarray}
\xi_ 1 &=& -\gamma _1 \lambda \kappa \nonumber\\
\xi_ 2 &=& -(\gamma_2+\gamma _3)\kappa \nonumber\\
\xi_ 3 &=&{\gamma _3}{\kappa} \nonumber\\
\xi_ 4 &=&(\gamma _2-\gamma _3)\kappa \nonumber\\
\xi_ 5 &=& \gamma _3 \kappa\nonumber\\ 
 \end{eqnarray}
Putting  these in the exponentiation of $S_J$ in Eq.(23) we get Jacobian contribution of FFBRST transformation.
 \begin{eqnarray}
S_J(\kappa =1)&= &-\gamma _ 1 \lambda B^2-[(\gamma _2+\gamma _3)B(\partial_\mu A^\mu)]+
 {\gamma _3} B(\partial_\mu A^\mu +g A_\mu A^\mu )\nonumber\\ &+&({\gamma _2}-{\gamma _3})\bar c\Box c +
{\gamma _3}{\bar c}[\Box +{2g}{A}^\mu{\partial_\mu}]c
\end{eqnarray}
This contribution of $S_J$ at $\kappa =1$, when adds to Lorenz  gauge  action provides effective t'Hooft Veltman action.
\begin{eqnarray} 
 S_{eff}^t=S^{L}+S_{J}(\kappa =1)
 \end{eqnarray}
\begin{eqnarray}
S_{eff}^t&=&\int d^4x[-\frac{1}{4}F^{\mu\nu}F_{\mu\nu}+\frac{\lambda }{ 2}(1-2\gamma _1) B^2-(\gamma _2-1)B(\partial_\mu A^\mu)
+(1+\gamma _2){\bar c\Box  c}\nonumber\\ &+&\gamma _3 B g A_\mu A^\mu
+{2\gamma _3}{\bar c}{A}^{\mu}{\partial_\mu} c ]\nonumber\\
\end{eqnarray}
This depicts the theory in t'Hooft  Veltman gauge with a different gauge fixing parameter\\

\begin{eqnarray}
   \zeta \longrightarrow \frac{\lambda }{2}(1-2\gamma _1)
\end{eqnarray}
Thus FFBRST with finite field dependent parameter given in Eq. (21) relate the generating functional corresponding to effective theories in Lorenz gauge and that of in t'Hooft Veltmann gauge.
\begin{eqnarray}
      Z_L \stackrel{FFBRST} \longrightarrow Z_t 
\end{eqnarray}
It means that  under FFBRST transformation the original  action in Lorenz gauge gets transformed into  effective action in t'Hooft gauge in path integral formulation.
\section {INTER-RELATION BETWEEN EFFECTIVE THEORIES IN GQED}
Now we would like to show that this generalized theory of Podolsky in different gauges are basically inter-related through field transformation.
This field transformation is nothing but the well known FFBRST transformation. For this purpose we construct the field dependent parameter $ \Theta= \int \Theta^{'}{d\kappa[\phi(x,\kappa)]}$ with 
\begin{eqnarray}
\Theta^{'}=i \int{d^4y}\bar c(y)[\gamma _ 1 \frac{\xi}{2} B+\gamma_ 2 (\frac{\Box}{{m_p}^2}+1)^{\epsilon}{\partial_\mu}{A}^{\mu}+ \gamma_3{\partial_\mu A ^\mu}]
\end{eqnarray}
The infinitesimal change in Jacobian due to such a finite transformation then can be calculated using Eq.(16 ) as,
\begin{eqnarray}
\frac {1}{J}\frac{dJ}{dk}&=&\int{dk}[(s_b{A}_\mu)\frac{\delta\Theta^{'}}{\delta A_\mu}-(s_b\bar c)\frac{\delta\Theta^{'}}{\delta\bar c}]\nonumber\\
&=& i \int{dk}[\gamma_ 2\bar c(\frac{\Box}{{m_p^2}}+1)^{\varepsilon } {\Box c}+\gamma_3 \bar c \Box c -\frac {\xi  \gamma _1}{2}B^2-\gamma_ 2 B  (\frac{\Box}{{m_p^2}}+1)^{\varepsilon}{\partial_\mu A^\mu} -\gamma_3 B\partial_\mu A^\mu ]\nonumber\\
\end{eqnarray}
We make an ansatz for the Jacobian contribution of this theory as,  
\begin{eqnarray}
S_J=\xi _1(k)B^2+\xi _2(k) B({\partial_\mu A^\mu})+\xi _3(k) B (\frac{\Box }{{m_p}^{2}}+1)^{\varepsilon }{\partial_\mu A^\mu}+\xi _4 (k){\bar c}(\frac{\Box} {{m_p}^{2}}+1)^{\varepsilon }\Box {c}+\xi _5(k)\bar c\Box {c}\nonumber\\
\end{eqnarray}
where $\xi _i(\kappa )$ are $\kappa $ dependent parameter with initial condition $ \xi _i(\kappa =0)=0$,
which will satisfy the condition in Eq.(15). Now we calculate
\begin{eqnarray}
\frac{dS_J}{dk}&=&[\xi_1^{'} B^{2}+\xi_2^{'} B(\partial_\mu A^\mu)+\xi_3^{'} B (\frac{\Box}{{m_p}^2}+1)^{\varepsilon }\partial_\mu A^\mu +\xi_4^{'} \bar c(\frac{\Box}{{m_p}^{2}}+1)^{\varepsilon }\Box c
+\xi_5^{'}\bar c\Box c\nonumber\\ &-& \xi _2  B \Box c \Theta^{'}-  \xi _3 B (\frac{\Box } {{m_p}^2}+1)^{\varepsilon }\Box  c\Theta ^{'} -\xi_ 4 B(\frac{\Box}{m_p^2}+1)^{\varepsilon }{\Box}c\Theta^{'} -{\xi_5}B{\Box}c\Theta^{'}]
\end{eqnarray}
Putting Eq.(34) and Eq.(36) in Eq.(15) we obtain,
\begin{eqnarray}
\int [&+&\gamma_ 2{\bar c}(\frac{\Box }{{m_p}^2}+1)^{\varepsilon } {\Box c}+\gamma_3 \bar c \Box c-{\frac{\xi}{2}}{\gamma _1}{B}^{2}-{\gamma_2(\frac{\Box}{{m_p}^2}+1})^{\varepsilon }B (\partial_\mu A^\mu) -\gamma_3 B (\partial_\mu A ^\mu) -{\xi _1}^{'}{B}^{2}\nonumber\\ &-&{\xi _2}^{'}{B}({\partial_\mu A^\mu})
-{\xi _3}^{'} B (\frac{\Box }{{m_p}^2}+1)^{\varepsilon }{\partial_\mu A^\mu}-{\xi _4}^{'}{\bar c}(\frac{\Box }{{m_p}^2}+1)^{\varepsilon }{\Box }{c}-{\xi _5}^{'}{\bar c}{\Box }{c}\nonumber\\ &+&{\xi _2}{B}{\Box }{c}\Theta ^{'}
+{\xi _3}{B}(\frac{\Box }{{m_p}^{2}}+1)^{\varepsilon }{\Box }{c}{\Theta }^{'}+{\xi _4}B(\frac{\Box }{{m_p}^2}+1)^{\varepsilon }{\Box }{c}{\Theta  }^{'}\nonumber\\&+&{\xi _5}{B}{\Box }{c}{\Theta }^{'}]. e^ {i (S_1+S_eff)}=0.
\end{eqnarray}
Now equating the coefficients of various terms in both side of the above equation we obtain,
\begin{eqnarray}
\xi _1^{'}+\frac{\xi }{2} \gamma _1(k)=0\nonumber\\
\xi_ 2^{'}+\gamma_ 3=0\nonumber\\
\xi_ 3^{'}+\gamma_ 2=0\nonumber\\
\xi _4^{'}-\gamma_2 = 0\nonumber\\
\xi_ 5^{'}-\gamma_ 3=0\nonumber\\
 \xi_ 2 +\xi_ 5=0\nonumber\\
 \xi_ 3+\xi_ 4=0\nonumber\\
\end{eqnarray}
Solving these equations subjected to the initial condition on $\xi_i(\kappa =0)=0$ we obtain, 
\begin{eqnarray}
\xi_ 1(k)=- \frac{\xi}{2} \gamma _1 k\nonumber\\
\xi_ 2(k)=  -\gamma_ 3 \kappa\nonumber\\
\xi_ 3(k)= -\gamma_ 2 \kappa\nonumber\\
\xi_ 4(k)= +\gamma_ 2 \kappa \nonumber\\
\xi_ 5(k)=  +\gamma_ 3 \kappa\nonumber\\
\end{eqnarray}
Thus we can say that a finite transformation of the type in Eq.(14) with the transformation parameter $ \Theta^{'}$ defined in Eq.(33) give rise to a local Jacobian factor of the type $\exp(iS_J)$ where,
\begin{eqnarray}
S_J (\kappa =1)={-}{\gamma _1}(k) {\frac{\xi}{2}} B^2-\gamma_ 3 B (\partial_\mu A^\mu)-\gamma_ 2 B  (\frac{\Box }{{m_p}^2}+1)^{\varepsilon }{\partial_\mu A^\mu}+\gamma_ 2\bar c(\frac{\Box }{{m_p}^{2}}+1)^{\varepsilon } {\Box c}+\gamma_ 3{\bar c}{\Box } c\nonumber\\
\end{eqnarray}
This contribution of $ S_J$ at $\kappa=1$, when adds to Lorenz gauge action provide effective generalized Lorenz gauge.
\begin{eqnarray}
\int D\phi e^{iS^{L}+{S_J}(\kappa =1)}=\int D\phi e^{iS^{P}}
\end{eqnarray}
effective action $S^P$ is given by
\begin{eqnarray}
S_{eff}^P&=&\int{d^4x}[- \frac{1}{4}F_{\mu\nu}F^{\mu\nu}+\frac{1}{2{m_p}^2} \partial_\mu F^{\mu\lambda}{\partial_\theta }F_{\theta }^{\lambda }+(1-\gamma_ 1) \frac{\xi }{2} B^2
+(1-\gamma_ 3) B \partial_\mu A ^\mu \nonumber\\ &+&(1+\gamma_ 3 ) \bar c \Box  c - \gamma_ 2 B(\frac{\Box}{{m_p}^2}+1)^\varepsilon \partial_\mu A^\mu +\gamma_ 2\bar c(\frac{\Box}{{m_p}^2}+1)^\varepsilon  \Box c]
\end{eqnarray}
Thus FFBRST in Eq.(33) takes $Z^L$ to $Z^P$.
\begin{eqnarray}
  Z^{L}\stackrel{FFBRST}\Longrightarrow Z^{P} 
\end{eqnarray}
with gauge parameter in the new theory is related to
\begin{eqnarray}
           \zeta \rightarrow (1-\gamma _1)
\end{eqnarray}
\section{Conclusion}
A generalized electromagnetic theory was proposed to resolve certain difficulties of Maxwell theory of electromagnetism by Podolsky. This theory contains higher order derivative terms and characterized by a free parameter Podolsky mass. The quantization for such theory requires  generalized Lorenz gauge condition $(1+\frac{\Box}{m_p^2})^\varepsilon \partial_\mu A^\mu=0$.  BRST invariant effective theories are developed for this theory for $\varepsilon =0$ (usual Lorenz gauge), $\varepsilon =\frac{1}{2}$ (no mixing gauge ). We have shown that  these effective theories are related through field transformation,which has been identified as finite field dependent  BRST transformation. The nontrivial  Jacobian for the Path integral measure is responsible for these 
connections. Thus the generating functionals corresponding to different BRST Invariant effective actions in the QED with higher derivative terms, as developed by Podolsky, are interrelated through appropriately constructed FFBRST transformation. Hence FFBRST transformation helps to construct theory in no-mixing gauge which is free from UV divergence in the radiative correction in the GQED. Our technique may be helpful in understanding the so called $"\frac{4}{3}$ problem" in Maxwell theory by connecting it to Podolsky theory.

\end{document}